\documentstyle[twocolumn,psfig]{mn}
\def\fr#1#2{\textstyle {#1\over #2}\displaystyle}

\title[Moving groups in the Hipparcos Catalogue]{
       A refurbished convergent point method for finding moving
       groups in the Hipparcos Catalogue\thanks{Based on data
       from the Hipparcos astrometry satellite.}}

\author[Jos H.J.\ de Bruijne]
       {Jos H.J.\ de Bruijne\\
        Sterrewacht Leiden, Postbus 9513, 2300 RA Leiden, the Netherlands}

\begin{document}

\maketitle

%%%%%%%%%%%%%%%
% Abstract
%%%%%%%%%%%%%%%

\begin{abstract}
The Hipparcos data allow a major step forward in the research of
`moving groups' in the Solar neighbourhood, as the common motion of
group members causes converging proper motions. Previous knowledge on
these coherent structures in velocity space has always been limited by
the availability, reliability, and accuracy of ground-based proper
motion measurements.

A refurbishment of Jones' convergent point method is presented which
takes full advantage of the quality of the Hipparcos data. The
original implementation of this method determines the maximum
likelihood convergent point on a grid on the sky and simultaneously
selects group members from a given set of stars with positions and
proper motions. The refurbished procedure takes into account the full
covariance matrix of the Hipparcos measurements instead of standard
errors only, allows for internal motions of the stars, and replaces
the grid-based approach by a direct minimization. The method is tested
on Monte Carlo simulations of moving groups, and applied to the
Hyades. Despite the limited amount of data used by the convergent
point method, the results for stars in and around the cluster-centre
region agree very well with those of the recent comprehensive study by
Perryman et al.
\end{abstract}

%%%%%%%%%%%%%%%
% Key words
%%%%%%%%%%%%%%%


%%%%%%%%%%%%%%%
% Main Text
%%%%%%%%%%%%%%%

%\section 1
\section{Introduction}
\label{sec_intro}

In the early years of this century, the acquisition of stellar
positions and proper motions was one of the main aims of astronomical
institutes. L.\ Boss' Preliminary General Catalogue (1910) revealed
`streams of stars' which showed converging proper motions (e.g.,
Kapteyn 1914, 1918; Eddington 1914). These streams were correctly
interpreted as groups of stars with a common space motion: `moving
groups'. As projection effects are responsible for the converging
patterns, proper motions can be used to establish membership of moving
groups which are sufficiently extended on the sky. Therefore,
convergent point studies have always been limited to nearby open
clusters such as the Hyades (e.g., van Bueren 1952) and OB
associations such as Scorpius OB2 (e.g., Blaauw 1946; Bertiau
1958). Even though associations are gravitationally unbound, they form
a coherent structure in velocity space: the initial velocity
dispersion of association members is a few km~s$^{-1}$ (e.g., Mathieu
1986), similar to that of the gas in the parental molecular cloud. Due
to expansion, tidal disruption by differential Galactic rotation, and
stripping by passing molecular clouds, the size of an association as
well as its velocity dispersion grow with time, until the group
eventually merges with the Galactic disk field star population.
Associations generally have large physical dimensions without clear
boundaries, and often show signs of substructure in the form of
subgroups (e.g., Blaauw 1964a), which may have different ages and
kinematics (e.g., Elmegreen \& Lada 1977).

In the pre-Hipparcos era, reliable high-accuracy proper motions were
available for a limited set of stars only. In order to avoid zonal
errors and other systematic effects, proper motion measurements were
restricted to small samples of bright stars in fundamental catalogues
or to small areas covered by a single photographic
plate. Consequently, astrometric membership of nearby associations,
which cover tens to hundreds of square degrees on the sky, has always
been limited to spectral types earlier than $\sim$B5, thus hiding
their low-mass stellar content. Consequently, many issues related to
the structure and dynamics of associations are still unknown (e.g.,
Blaauw 1964a, 1991). A well-defined membership selection is the
necessary first step to solve these open questions. Based on
Hipparcos\footnote{
The Hipparcos Catalogue (ESA 1997) is an all-sky astrometric catalogue
containing absolute positions, proper motions, and parallaxes for
118,218 selected stars with accuracies of $\sim$1~${\rm mas}~({\rm
yr}^{-1})$. The limiting magnitude is $V \sim 12.4$~mag, and the
completeness limit is $V \sim 7.3$--9.0~mag, depending on spectral
type and position on the sky. The Catalogue also contains duplicity,
photometric, and variability information.}
data, de Zeeuw et al.\ (1999) have carried out a census of the stellar
content of the nearby associations. Membership was established based
on two independent selection methods. Both are based on the concept of
common space motion: the `Spaghetti method' uses proper motions {\it
and$\,$} parallaxes to search for the common motion of stars in
velocity space (Hoogerwerf \& Aguilar 1999); the convergent point
method is discussed in this paper. It is based on proper motion data
only.

Early algorithms implementing the convergent point method are due to
Charlier and Bohlin (1916; \S \ref{subsec_Charlier}; cf.\ Smart
1968). Our version is basically a refurbishment of Jones' (1971)
method (\S \ref{subsec_clas}). Because of the limited computer power
in the early 70's, Jones' original implementation determined the
maximum likelihood convergent point on a grid on the sky. We have
replaced the grid-approach by a direct minimization routine. We also
take into account the internal motions of the group members, as well
as the full Hipparcos covariance matrix (\S \ref{subsec_mod}).

The convergent point method selects members based on proper motions
and positions, but does not use any assumption about their
distribution in configuration space. Therefore, the method can detect
moving groups `pur sang', the members of which fill a small volume in
velocity space. However, any attempt to find such moving groups in the
{\it full$\,$} Hipparcos Catalogue using proper motion data only is
likely to yield a non-physical detection, as any convergent point has
an associated `moving group' covering large parts of the sky (\S
\ref{subsec_appl_super}). In such cases, inclusion of parallax, radial
velocity, and/or photometric data is essential to decide whether the
{\it space velocities$\,$} of the candidate members are identical or
not. Therefore, we have limited ourselves in this paper to test and
apply the convergent point method to open clusters and
associations. These two special types of moving groups (members of an
open cluster or OB association have a common space motion as well as
configuration space position: they fill a restricted volume in phase
space) allow the use of finite-size fields on the sky, resulting in an
acceptable contamination by field stars (\S \ref{sec_test}).  Our
simulated groups have 30~pc diameters, requiring up to
$20^\circ\!$$\times$$20^\circ\!$ fields on the sky, and thus resemble
moving groups/associations rather than classical open clusters.

This paper is organized as follows. \S \ref{sec_ccpm} describes the
original and new implementation of the convergent point method. \S
\ref{sec_test} deals with the optimization of the algorithm based on
synthetic data, while \S \ref{sec_appl} presents an application to
the Hyades with Hipparcos data. \S \ref{sec_disc} discusses our
results.

%\section 2
\section{Convergent point method}
\label{sec_ccpm}

%\section 2.1
\subsection{Basic notation and definitions}
\label{subsec_basics}

The Hipparcos Catalogue (ESA 1997) provides stellar positions
$\alpha$, $\delta$ and proper motions $\mu_{\alpha} \cos\delta$,
$\mu_{\delta}$ (in ${\rm mas}~{\rm yr}^{-1}$) in the equatorial
coordinate system. Galactic coordinates are denoted by $\ell$, $b$
(ESA 1997, Vol.\ 1, \S 1.5). The trigonometric parallax $\pi$ (in mas)
is related to distance $D$ (in pc) according to: $D \equiv 1000 /
\pi$. The proper motions $\mu_\alpha \cos\delta$ and $\mu_\delta$ are
the components of the proper motion vector $\bmath{\mu}$ which
corresponds to the velocity tangential to the line of sight
$\bmath{v}_{\rm tan}$ (in ${\rm km}~{\rm s}^{-1}$) according to
$\bmath{v}_{\rm tan} \equiv A \bmath{\mu} / \pi$, where $A \equiv
4.740470446~{\rm km}~{\rm yr}~{\rm s}^{-1}$ is the ratio of one
astronomical unit in kilometers and the number of seconds in one
Julian year (ESA 1997, Vol.\ 1, table~1.2.2). For a given radial
velocity $v_{\rm rad}$ (in ${\rm km}~{\rm s}^{-1}$), the space
velocity $\bmath{v}$ (relative to the Sun, in ${\rm km}~{\rm s}^{-1}$)
then follows from:
\begin{equation}
\bmath{v} \equiv \left(
\begin{array}{c}
v_\alpha \\
v_\delta \\
v_{\rm rad}
\end{array} \right) \equiv
\left(\begin{array}{c}
A \mu_\alpha \cos\delta / \pi\\
A \mu_\delta            / \pi\\
v_{\rm rad}
\end{array}\right).
\label{eq_def_v}
\end{equation}
We denote by $(X,Y,Z)$ the linear velocity components corresponding to
$\bmath{v}$ with respect to the usual rectangular equatorial
coordinate system in which the $X$-component is directed towards the
vernal equinox, the $Y$-component is directed towards the point on the
equator with $\alpha = 90^\circ\!$, and the $Z$-component is directed
towards the northern equatorial pole.

In the absence of measurement errors and internal motions (velocity
dispersion), the proper motions of a set of stars with the same space
motion $\bmath{v}$ (a `moving group') converge on the sky:
\begin{equation}
\mu \equiv | \bmath{\mu} | = {{\pi v \sin\lambda}\over{A}},
\label{eq_def_mu_ideal}
\end{equation}
where $v \equiv | \bmath{v} |$, and 
\begin{equation}
\cos\lambda \equiv
\sin\delta \sin\delta_{\rm cp} +
\cos\delta \cos\delta_{\rm cp} \cos(\alpha_{\rm cp} - \alpha),
\label{eq_def_coslambda}
\end{equation}
and $\lambda$ is the angular distance between the position $(\alpha,
\delta)$ of any star and the position $(\alpha_{\rm cp}, \delta_{\rm
cp})$ of the convergent point. The convergent point denotes the
direction of the projected space velocity $\bmath{v}$ (\S
\ref{subsec_Charlier}). The radial velocities of the stars are given
by $v_{\rm rad} = v \cos\lambda$.

By transforming the proper motion components $\mu_\alpha \cos\delta$
and $\mu_\delta$ into components $\mu_{\parallel}$, directed parallel to
the great circle which joins any star and the convergent point, and
$\mu_{\perp}$, directed perpendicular to the same great circle, the
convergence of proper motions (eq.~\ref{eq_def_mu_ideal}) can be
expressed as:
\begin{equation}
\left\{
\begin{array}{rcl}
\mu_{\parallel} & = & \pi v \sin\lambda / A \ = \ \mu, \\ 
\mu_{\perp}     & = & 0,
\end{array}
\right.
\label{eq_perpar}
\end{equation}
with
\begin{equation}
\left(\begin{array}{c}
\mu_{\parallel} \\
\mu_{\perp}
\end{array}\right)
\equiv
\left(\begin{array}{rr}
 \sin\theta & \cos\theta \\
-\cos\theta & \sin\theta
\end{array}\right)
\left(\begin{array}{c}
\mu_\alpha \cos\delta \\
\mu_\delta
\end{array}\right),
\label{eq_def_perpar}
\end{equation}
where
\begin{equation}
\tan\theta = {{\sin(\alpha_{\rm cp} - \alpha)} \over
{\cos\delta \tan\delta_{\rm cp} - \sin\delta \cos(\alpha_{\rm cp} -
\alpha)}}.
\label{eq_def_theta}
\end{equation}
Thus, the expectation value of $\mu_{\perp}$ for any star is zero.

The $\mu_{\perp}$-distribution contains information on the velocity
dispersion of the stars and on the measurement errors. The
$\mu_{\parallel}$-distribution contains, besides these two effects, a
systematic perspective effect, depending on the angular distance
between the stars and the convergent point ($\sin\lambda$), and the
individual parallaxes ($\pi$).  Conversely, the assumption of a common
space motion $\bmath{v}$ for all stars allows the components
$\mu_{\parallel}$ to be used to {\it derive$\,$} individual parallaxes
(e.g., Jones 1971). A sophisticated method to obtain these so-called
secular parallaxes for a given set of moving group members is
described by Dravins et al.\ (1997).

Eq.~(\ref{eq_def_mu_ideal}), and thus eq.~(\ref{eq_perpar}), is also
valid if the stars are in a state of linear expansion (with expansion
coefficient $k$ in ${\rm km}~{\rm s}^{-1}~{\rm pc}^{-1}$; e.g., Blaauw
1956, 1964b). In this case, however, the factor $v$ in
eq.~(\ref{eq_def_mu_ideal}) no longer solely corresponds to the space
motion $\bmath{v}$ of the stars, but also depends on the expansion
velocity, i.e., $v$ should be replaced by $v^\prime = v^\prime (v,
k)$. As a result, a kinematical detection of expansion requires radial
velocities. It follows that if proper motions are used to determine
the space motion or convergent point of the stars, one derives the
genuine space motion {\it plus$\,$} the reflex of an expanding motion.

%\section 2.2
\subsection{Charlier's method}
\label{subsec_Charlier}

Consider a set of stars with the same space motion $\bmath{v}$. For
each star, the 3 equations of condition are (e.g., Smart 1968):
\begin{equation}
\left(\begin{array}{rrr}
-\sin\alpha\phantom{\cos\delta} &  \cos\alpha\phantom{\sin\delta}& 0          \\
-\cos\alpha         \sin\delta  & -\sin\alpha         \sin\delta & \cos\delta \\
 \cos\alpha         \cos\delta  &  \sin\alpha         \cos\delta & \sin\delta
\end{array}\right)
\left(\begin{array}{c}
X\\
Y\\
Z
\end{array}\right)
=
\bmath{v}.
\label{eq_eq_of_cond}
\end{equation}
The 3$\times$3 matrix and the vector $\bmath{v}$ (eq.~\ref{eq_def_v})
follow directly from the observables $\alpha, \delta, \pi, \mu_\alpha
\cos\delta, \mu_\delta, v_{\rm rad}$. The space motion components
$(X,Y,Z)$, the modulus $v$ of the space motion vector $\bmath{v}$, and
the coordinates $(\alpha_{\rm cp}, \delta_{\rm cp})$ of the convergent
point are related through:
\begin{equation}
\left(\begin{array}{c}
X\\
Y\\
Z
\end{array}\right)
\equiv v
\left(\begin{array}{r}
\cos\alpha_{\rm cp} \cos\delta_{\rm cp}\\
\sin\alpha_{\rm cp} \cos\delta_{\rm cp}\\
                    \sin\delta_{\rm cp}
\end{array}\right).
\label{eq_def_XYZ}
\end{equation}
The parallax $\pi$ can be eliminated from the first two lines of
eq.~(\ref{eq_eq_of_cond}) by use of eq.~(\ref{eq_def_v}), which leads
to the equation:
\begin{equation}
a X + b Y + c Z = 0,
\label{eq_Charlier}
\end{equation}
where the coefficients $a$, $b$, and $c$ depend on the observables
$\alpha, \delta, \mu_\alpha \cos\delta, \mu_\delta$. The resulting
equation of condition (\ref{eq_Charlier}) can be solved for a set of
stars in a least-squares sense to yield the ratios $X:Y:Z$, and thus
the convergent point:
\begin{equation}
\alpha_{\rm cp} = \arctan\left(Y   \over X                 \right), \qquad
\delta_{\rm cp} = \arctan\left({{Z}\over{\sqrt{X^2 + Y^2}}}\right).
\end{equation}
The third element of the space motion, its modulus $v$, can be
determined only if radial velocities are available. This solution
procedure is due to Charlier\footnote{
Independently, Bohlin arrived at an equation of the form $a^{\prime} X
+ b^{\prime} Y + c^{\prime} Z = 0$. However, Bohlin's and Charlier's
equations are equivalent, except for different weights of the
coefficients: $a^{\prime} = a / \mu$, $b^{\prime} = b / \mu$, and
$c^{\prime}=c / \mu$. See Smart (1968).}
(1916). However, Charlier's method incorrectly treats the
least-squares coefficients $a$, $b$, and $c$, which depend on
observables with measurement errors, as {\it constants}. This leads to
systematic errors, which were sometimes `corrected' for differentially
(e.g., Seares 1944, 1945; Petrie 1949a, b; Roman 1949).

%\section 2.3
\subsection{Jones' method}
\label{subsec_clas}

Based on the work by Brown (1950), Jones (1971) developed a maximum
likelihood method for a simultaneous determination of convergent point
and moving group membership. This procedure avoids the conceptual
problems of Charlier's method (\S \ref{subsec_Charlier}). The basic
idea is to determine the maximum likelihood coordinates $(\alpha_{\rm
cp}, \delta_{\rm cp})$ of the convergent point based on a comparison
of the proper motion components $\mu_{\perp}$ with their expectation
value of zero (\S \ref{subsec_basics}).

Jones starts with a sample of $N$ stars with known positions, proper
motions, and corresponding errors. He overlays the sky $\alpha =
0^\circ\!$--$360^\circ\!$, $\delta = -90^\circ\!$--$90^\circ\!$ with a
grid of trial convergent points (e.g., 24$\times$12 cells of
15$^\circ\!$$\times$15$^\circ\!$), and numbers each grid point (e.g.,
$i = 1, \ldots, N_{\rm grid} = 288$). Then, he follows the recipe
outlined below.
\begin{enumerate}
\item Start at grid point $i = 1$.
\item Assume that this grid point {\it is$\,$} the convergent
point. The coordinates of this point are referred to as $(\alpha_{\rm
cp}, \delta_{\rm cp})$.
\item Calculate for each star $j$ the error-weighted value $t_\perp$
of the component $\mu_\perp$ (eq.~\ref{eq_def_perpar}):
\begin{equation}
t_\perp \equiv {{\mu_{\perp}}\over{\sigma_\perp}},
\label{eq_def_t}
\end{equation}
where to first order:
\begin{equation}
\sigma_\perp^2 = (\sigma_{\theta}           \mu_{\parallel})^2 +
                 (\sigma_{\mu_\alpha \cos\delta} \cos\theta)^2 +
                 (\sigma_{\mu_\delta}            \sin\theta)^2,
\label{eq_def_sigperp}
\end{equation}
and $\sigma_\theta$ follows from eq.~(\ref{eq_def_theta}). Jones
assumes that the quantity $t_{\perp}$ is distributed normally with
zero mean and unit variance. Thus, the probability distribution for
the given combination of $\mu_{\perp,j}$ and $\sigma_{\perp,j}$ to
occur is given by:
\begin{equation}
p_j \equiv {{1}\over{\sqrt{2\pi}}} \exp{(-\fr{1}{2} t_{\perp,j}^{2})}.
\label{eq_def_p_j}
\end{equation}
For a set of stars with the same space motion, each proper motion
vector transforms into the component $\mu_{\parallel}$; consequently,
the expectation value for $\mu_{\perp}$ equals zero
(eq.~\ref{eq_perpar}).
\item Evaluate at the grid point the quantity $X^2$:
\begin{equation}
X^2 \equiv \sum_{j = 1}^{N} t_{\perp,j}^{2}.
\label{eq_def_chi}
\end{equation}
As $t_\perp$ is distributed normally (cf.\ step~[iii]), $X^2$ is
distributed as $\chi^2$ with $N - 2$ degrees of freedom.
\item Determine $X^2$ for each grid point: repeat steps~(i) through
(iv) for $i = 1, \ldots, N_{\rm grid}$.
\item The total probability $P$ for the given set of calculated
values of $t_\perp$ to occur is given by:
\begin{equation}
P \equiv \prod_{j=1}^{N} p_j
  =      {{1}\over{(2\pi)^{N/2}}} \exp{(-\fr{1}{2}X^2)}.
\label{eq_def_ptot}
\end{equation}
Thus, the likelihood function ${\cal L}$ can be defined as:
\begin{equation}
{\cal L} \equiv {(2\pi)^{N/2}} P
         =      \exp{(-\fr{1}{2}X^2)},
\label{eq_def_L}
\end{equation}
and maximizing ${\cal L}$ is equivalent to minimizing
$X^2$. Therefore, the grid point with the lowest value of $X^2$ is
defined as the convergent point $(\alpha_{\rm cp}, \delta_{\rm
cp})$.
\item Given the number of degrees of freedom, evaluate the
probability $\epsilon$ that $X^2$ will exceed the observed value of
$X^2$ by chance {\it even$\,$} for a correct model (e.g., Press et
al.\ 1992):
\begin{equation}
\epsilon \equiv {{1}\over{\Gamma({{1}\over{2}}[N-2])}}
   \int_{X^2}^{\infty} {\rm d}x\ x^{{{1}\over{2}}[N-2] - 1} \exp{(-x)},
\label{eq_def_eps}
\end{equation}
where $\Gamma(x)$ denotes the Gamma function for $x > 0$.
\item If the computed value of $X^2$ is unacceptably high ($\epsilon <
\epsilon_{\rm min}$), reject the star with the highest value of
$|t_{\perp}|$, correct $N$$\rightarrow$$N - 1$, and go to step~(i). If
the value of $X^2$ is acceptable ($\epsilon \geq \epsilon_{\rm min}$),
go to step~(ix).
\item The maximum likelihood convergent point is chosen as the grid
point $(\alpha_{\rm cp}, \delta_{\rm cp})$, and all non-rejected stars
in the sample are identified as members (cf.\ \S
\ref{subsec_disc}). Thus, the determination of the moving group
convergent point and membership are intricately linked.
\end{enumerate}

It is important to reject in advance from the sample all stars with
insignificant proper motions (cf.\ Jones 1971). Insignificant in this
respect means:
\begin{equation}
t \equiv {{\mu} \over {\sigma_\mu}}
\equiv {{\sqrt{{\mu_{\alpha}^{2} \cos^{2}\delta} + \mu_\delta^{2}}}\over
       {\sqrt{\sigma_{{\mu_\alpha \cos\delta}}^{2} +
              \sigma_{{\mu_\delta           }}^{2}}}}
\leq t_{\rm min}.
\label{eq_def_tpm}
\end{equation}
Because such stars have proper motions that carry `no information',
the corresponding values for $t_{\perp}$ are small, independent of the
position of the convergent point. Thus, these stars are likely never
to be rejected.

The convergent point method selects as members all stars with proper
motion components $\mu_\perp$ that are close to their expectation
value of zero. However, not all stars with a small $\mu_\perp$
necessarily are moving group members: stars at large distances
generally have small proper motions, and correspondingly also small
components $\mu_\perp$. Thus, the convergent point membership
selection is biased towards distant stars (cf.\ \S
\ref{subsec_appl_open}).

%\section 2.4
\subsection{A refurbished convergent point method}
\label{subsec_mod}

In view of the release of the Hipparcos data, we have modified the
convergent point method described in \S \ref{subsec_clas} in three
ways, and extended it to include membership probabilities.

%\section 2.4.1
\subsubsection{Error propagation}
\label{subsubsec_mod1}

Unlike previous astrometric catalogues which quote (1$\sigma$)
standard errors for the observables, the Hipparcos Catalogue provides
the full covariance matrix for the measured astrometric parameters.
Thus, the propagation of errors (e.g., eq.~\ref{eq_def_sigperp}) must
take the full covariance matrix into account. A general prescription
to do so is given by ESA (1997, Vol.\ 1, \S 1.5). The specific case
for the transformation of $(\mu_\alpha \cos\delta, \mu_\delta)$ to
$(\mu_{\parallel}, \mu_{\perp})$ is described in
Appendix~\ref{app_mu_trans}.

%\section 2.4.2
\subsubsection{Internal motions}
\label{subsubsec_mod2}

With infinitely accurate measurements, moving group members would not
necessarily have proper motions directed exactly towards the
convergent point because of their velocity dispersion. Therefore,
selecting stars with $t_{\perp} = 0$ will not identify all
members. Consequently, when dealing with accurate proper motions, a
certain amount of deviation of $t_{\perp}$ from 0 should be allowed in
the selection procedure. This is achieved by changing the definition
of $t_{\perp}$ (eq.~\ref{eq_def_t}) to:
\begin{equation}
t_{\perp} \equiv {{\mu_{\perp}}\over{\sqrt{\sigma_\perp^2 +
                                          {\sigma_{\rm int}^\star}^2}}},
\label{eq_def_tnew}
\end{equation}
where $\sigma_{\rm int}^\star$ is an estimate of the one-dimensional
velocity dispersion in the group, in proper-motion units. If
$\sigma_{\rm int}$ is the one-dimensional velocity dispersion (in
km~s$^{-1}$) and $D$ is the distance of the group (in pc) one has:
\begin{equation}
\sigma_{\rm int} = A \sigma_{\rm int}^\star D / 1000.
\label{eq_def_sigma}
\end{equation}
The definition of $t$ (eq.~\ref{eq_def_tpm}) is also modified:
\begin{equation}
t \equiv {{\mu} \over {\sqrt{{\sigma_\mu}^2 + {\sigma_{\rm int}^\star}^2 }}}.
\label{eq_def_tpmnew}
\end{equation}

Jones acknowledges that the assumption of a normal
$t_{\perp}$-distribution is critical to the use of the $\chi^2$
distribution, but states that `... the rejection of individual stars
depends on the much weaker assumption that $p_j$ decreases
monotonically as $t_\perp$ increases'. We have carried out Monte Carlo
simulations of moving groups with different distances, physical sizes,
and velocity dispersions, in order to assess whether the observed
$t_\perp$-distributions are normal with zero mean and unit variance.
Kolmogorov--Smirnov tests reveal no significant differences between
the simulations and the model, independent of group distance, size,
and velocity dispersion, though the inclusion of internal motions,
i.e., usage of eq.\ (\ref{eq_def_tnew}) instead of (\ref{eq_def_t}),
is essential.

%\section 2.4.3
\subsubsection{Direct minimization}
\label{subsubsec_mod3}

The original approach of evaluating $X^2$ on a grid (\S
\ref{subsec_clas}) is limited in the sense that the accuracy of the
position of the convergent point is restricted by the mesh size of the
grid. Although a zoom-in strategy could overcome this problem, we can
now drop the grid-based approach, i.e., improve the numerical
implementation of Jones' method. In order to find the maximum
likelihood convergent point, we apply a global direct minimization
routine in two dimensions. This routine explicitly uses the
analytically known derivatives $\partial X^2 / \partial \alpha_{\rm
cp}$ and $\partial X^2 / \partial \delta_{\rm cp}$ of the objective
function $X^2$ with respect to the free variables $\alpha_{\rm cp}$
and $\delta_{\rm cp}$ (eq.~\ref{eq_def_chi}). It returns the
convergent point, as well as its uncertainties in the form of a
2$\times$2 covariance matrix.

%\section 2.4.4
\subsubsection{Membership probabilities}
\label{subsubsec_mod4}

A natural membership probability should be defined in the
$\mu_\parallel$--$\mu_\perp$-plane, as our membership selection takes
place in this plane. For any star, the relevant observables are the
proper motion components $\mu_\parallel$ and $\mu_\perp$ and the
elements of the corresponding 2$\times$2 covariance matrix ${\bf C}$
(Appendix~\ref{app_prob}, eq.~\ref{def_eq_C2x2}), which describes the
shape and orientation of the confidence region related to the
Hipparcos measurements. A membership probability $p$ can be defined as
$p \equiv 1 - p_{\rm conf}$, where $p_{\rm conf}$ is the {\it
minimum$\,$} confidence limit of the confidence ellipse that is
consistent with $\mu_{\perp} = 0$. In Appendix~\ref{app_prob}, it is
shown that $p_{\rm conf} = 1 - \exp{(-\fr{1}{2}[{{\mu_\perp} /
{\sigma_\perp}}]^2)}$, so that the membership probability $p$ for a
star is:
\begin{equation}
p = \exp{(-\fr{1}{2}\left[{{\mu_\perp}\over{\sigma_\perp}}\right]^2)}.
\label{eq_def_pconf}
\end{equation}
In view of the new definitions for $t_{\perp}$ and $t$
(eqs~\ref{eq_def_tnew} and \ref{eq_def_tpmnew}) which take into
account the velocity dispersion of the group, we modify the membership
probability definition to:
\begin{equation}
p = \exp{(-{{1}\over{2}}\left[{{\mu_\perp^2}\over{
          \sigma_\perp^{2} +{ \sigma_{\rm int}^\star}^2
        }}\right])}.
\label{eq_def_pnew}
\end{equation}

%\section 2.5
\subsection{Discussion}
\label{subsec_disc}

Whether a specific star is rejected or not depends, apart from
$(\alpha_{\rm cp}, \delta_{\rm cp})$, not only on its individual
membership probability but also on the probabilities $p_j$ of all
other stars, since the product of the $p_j$ is considered
(eqs~\ref{eq_def_ptot} and \ref{eq_def_L}). More specifically, if
there are many stars with high $p_j$ (low $t_\perp$; very well
converging proper motions) more low-$p_j$ (high-$t_\perp$) stars can
be tolerated. In other words: a moving group with a small velocity
dispersion allows the method to include some non-members. However,
this effect cannot be large as the velocity dispersion of the group is
accounted for in the model (\S \ref{subsubsec_mod2}).

Instead of using $t_\perp$ as membership indicator one could introduce
a polar angle $\phi_t$ in the $t$-plane ($t_\parallel \equiv t
\sin\phi_t$, $t_\perp \equiv t \cos\phi_t$), and consider its
distribution instead of $f(t_\perp)$ (see Appendix~\ref{app_alt} for
details). Tests indicate that this alternative membership
determination procedure gives very similar results to the procedure
presented in \S\S \ref{subsec_clas}--\ref{subsec_mod}.

The current numerical implementation of the refurbished convergent
point method is not able to cope with samples containing two or more
moving groups. The procedure selects stars which are related to the
lowest valley (measured in terms of $X^2$) in $(\alpha_{\rm cp},
\delta_{\rm cp})$-country. A detection of all minima, each one
indicative of an over-abundance of high-$p_j$ stars, could
simultaneously identify multiple moving groups. The statistical
significance of each minimum could be assessed by considering the
$p_j$-distribution and comparing it with the expectation from field
stars only. This method would be the two-dimensional analogue of
Hoogerwerf \& Aguilar's (1999) three-dimensional Spaghetti method.

%\section 3
\section{Optimization of the algorithm}
\label{sec_test}

\begin{figure}
\centerline{\psfig{file=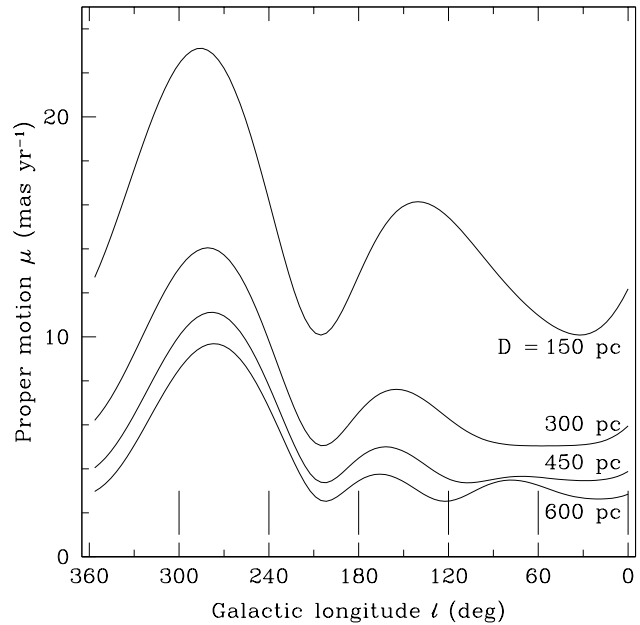,width=8.35cm,silent=}}
\caption{The effect of Galactic rotation and Solar motion on proper
motion $\mu$ as function of Galactic longitude $\ell$, at $b =
0^\circ\!$, for four different distances ($D = 150, 300, 450$, and
600~pc). The six vertical lines denote the specific choices for $\ell$
of the cluster centres in the synthetic data.}
\label{fig1}
\end{figure}

\begin{figure*}
\centerline{\psfig{file=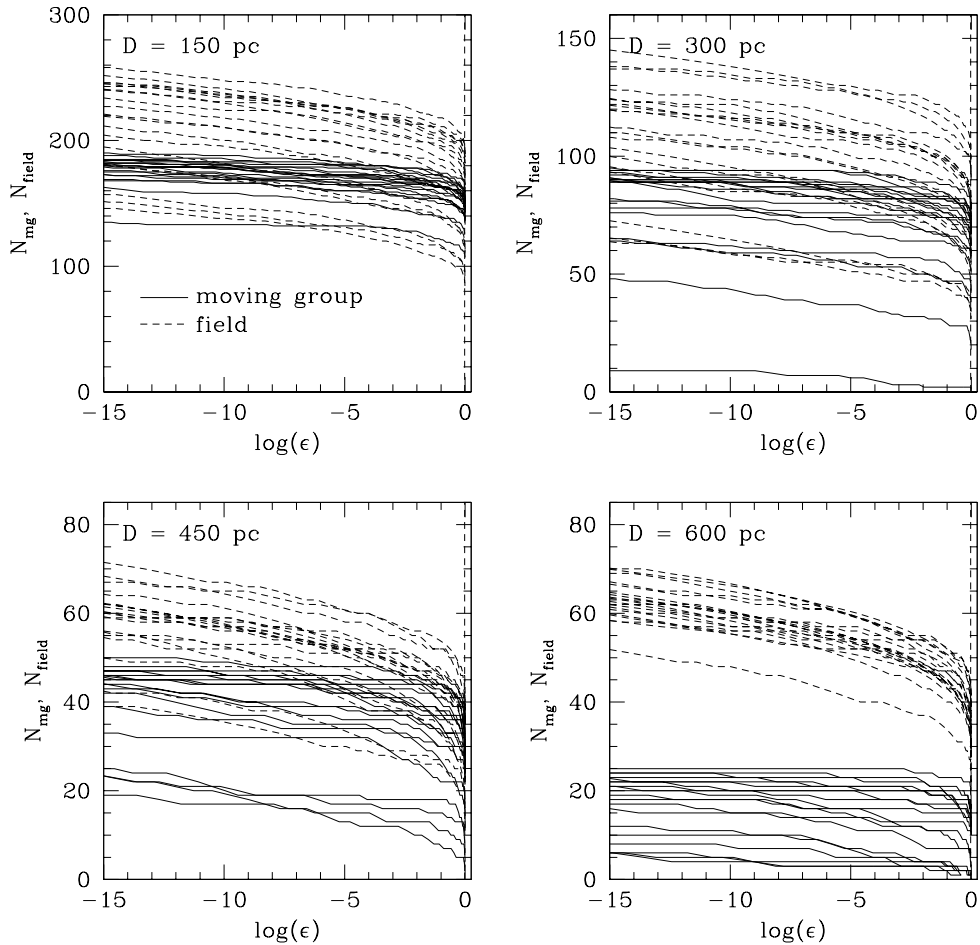,width=13.0truecm,silent=}}
\caption{Results of the membership selection procedure applied to
synthetic data with a cluster centred at $(\ell, b) = (300^\circ\!,
0^\circ\!)$ at a distance $D = 150, 300, 450$, and 600~pc. The panels
refer to the different cluster distances as indicated. Each panel
shows the number of accepted cluster stars (solid) and field stars
(dashed) as function of $\log\epsilon$ (eq.~\ref{eq_def_eps}). The
dashed vertical line at the right-hand side of each panel denotes
$\epsilon_{\rm min} = 0.954$ or $\log\epsilon_{\rm min} = -0.020$.
Note the different vertical scales of the panels.}
\label{fig2}
\end{figure*}

The convergent point method has several free parameters. In order to
understand their importance, and determine their optimum values (\S
\ref{subsubsec_appl}), we apply the method to synthetic data of
clusters superimposed on a kinematic model of the Galactic disk (\S
\ref{subsubsec_constr}). We have included the basic characteristics of
the Hipparcos Catalogue in the synthetic data.

%\section 3.1
\subsection{Construction of synthetic data}
\label{subsubsec_constr}

The synthetic data set consists of 24 distinct configurations: we vary
the Galactic longitude $\ell$ of the cluster centre from $\ell =
0^\circ\!$ to $300^\circ\!$, in steps of $60^\circ\!$; we assume a
constant Galactic latitude $b = 0^\circ\!$. We vary the distance of
the cluster centre from $D = 150$ to 600~pc, in steps of 150~pc. We
represent each synthetic cluster by $N_{\rm mg}$ stars distributed in
a sphere of radius 15~pc with a constant volume density. As the
cluster distance increases, the fictitious number of members observed
by Hipparcos decreases due to the completeness limit of the
Catalogue. Because we consider clusters with a fixed intrinsic size,
we decrease the size of the field of view with increasing cluster
distance. The number of field stars $N_{\rm field}$ is chosen such
that the total number of stars in the field is consistent with the
Hipparcos Catalogue mean stellar density of 3 stars per square
degree. The size of the field of view as well as the number of cluster
members $N_{\rm mg}$ have been chosen based on results obtained for
the nearby associations by de Zeeuw et al.\ (1999).
Table~\ref{tab_synth_data} summarizes the properties of the different
configurations.

\begin{table}
\caption{Characteristics of the synthetic data. The radius of the
spherical cluster is 15~pc, independent of its distance $D$. The total
number of stars $N_{\rm mg} + N_{\rm field}$ in each field of view
equals its size (Field size $\ell$$\times$$b$) times 3. The latter
number represents the mean stellar density per square degree in the
Hipparcos Catalogue.}
\label{tab_synth_data}
\begin{center}
\begin{tabular}{cccc}
\hline
\smallskip
$D$ (pc)                               &
Field size $\ell$$\times$$b$ (deg$^2$) &
$N_{\rm mg}$                           &
$N_{\rm field}$                        \\
150 & 20$\times$20 & 200 & 1000 \\
300 & 15$\times$15 & 100 &  575 \\
450 & 10$\times$10 &  50 &  250 \\
600 & 10$\times$10 &  25 &  275 \\
\hline\\ 
\end{tabular}
\end{center}
\end{table}

%\section 3.1.1
\subsubsection{Cluster kinematics}
\label{subsubsubsec_mg_kinematics}

The measured proper motion of each cluster star can be decomposed into
(1) a systemic streaming motion of the cluster with respect to its own
standard of rest, (2) a component due to velocity dispersion, (3) the
reflex of the Solar motion with respect to the Local Standard of Rest
(LSR), and (4) the effect of differential Galactic rotation. First, we
take effects (1)--(3) and observational errors into account; then we
add Galactic rotation to the proper motion.

Ad (1): the streaming velocity of the cluster is randomly drawn from a
constant-density sphere in velocity space with a radius of $10~{\rm
km}~{\rm s}^{-1}$. Ad (2): the velocity dispersion is assumed to be
isotropic, and each of the three independent components is drawn from
a Gaussian distribution with $\sigma = 2.0~{\rm km}~{\rm s}^{-1}$. Ad
(3): we adopt for the Solar motion with respect to the LSR the
Hipparcos value of Dehnen \& Binney (1998): $(U, V, W)_{\sun} =
(10.00, 5.23, 7.17)~{\rm km}~{\rm s}^{-1}$.

We translate the synthetic three-dimensional velocities and positions
into observables, after which observational errors are added: the
positions are assumed to remain unchanged, while we disturb the
parallax and proper motion by amounts which are randomly drawn from
Gaussian distributions with $\sigma = 1~{\rm mas}~({\rm yr}^{-1})$,
consistent with the median errors of the Hipparcos data. Then, we add
Galactic rotation to the proper motion (item 4) using the first-order
formulae (e.g., Smart 1968). We adopt for the Oort constants the
Hipparcos values of Feast \& Whitelock (1997): $A = 14.82$, $B =
-12.37~{\rm km}~{{\rm s}^{-1}}~{\rm kpc}^{-1}$.  Finally, the
covariance matrix elements for the synthetic member are directly
copied from a star which is randomly drawn from the subset of stars in
the Hipparcos Catalogue which have positive parallaxes.

%\section 3.1.2
\subsubsection{Field kinematics}
\label{subsubsubsec_field_kinematics}

The synthetic field star distribution is consistent with the
characteristics of the Hipparcos Catalogue as well as with the
kinematical properties of the local Galactic disk. We describe the
velocity distribution in the Solar neighbourhood in terms of the
Schwarzschild velocity ellipsoid (e.g., Binney \& Merrifield 1998).
This model is simple but adequate for our purposes. Based on Hipparcos
data, Dehnen \& Binney (1998) measured the first two moments of the
velocity distribution in the Solar neighbourhood as function of $B-V$
colour, as well as the Solar motion with respect to the LSR (\S
\ref{subsubsubsec_mg_kinematics}). We use their results to
characterize the shape and orientation of the velocity ellipsoid.

We randomly draw the required number of field stars from the subset of
stars in the Hipparcos Catalogue which have coordinates in the
corresponding field of view as well as positive parallaxes. We retain
all Hipparcos measurements, but replace the proper motions by
synthetic values which are projections\footnote{
The projection requires $\pi > 0$. Although other, more complicated,
methods could be used which do not require $\pi > 0$, it is not a
priori clear that such efforts give more realistic results, especially
because only 4~per cent of all Hipparcos stars has $\pi \leq 0$ (or no
parallax at all).}
of space velocities randomly drawn from Schwarzschild's model, given
the corresponding $B - V$ colour. Galactic rotation is added to the
proper motions as described in \S \ref{subsubsubsec_mg_kinematics}.

Following this procedure, the characteristics of the synthetic field
stars are consistent with the characteristics of the Hipparcos
Catalogue (e.g., the $V$ magnitude, parallax, and spectral type
distributions) as well as with the kinematics of the Galactic disk in
the corresponding field of view. De Zeeuw et al.\ (1999; their \S\S
3.4.3--3.4.5) extensively discuss Dehnen \& Binney's results and the
procedure described here in relation to similar Monte Carlo
simulations.

\begin{figure}
\centerline{\psfig{file=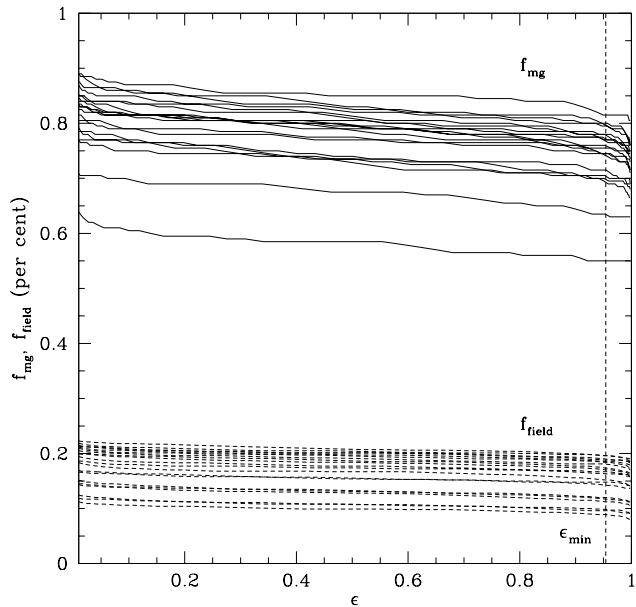,width=8.35cm,silent=}}
\caption{The fractions $f_{\rm mg}$ (solid) and $f_{\rm field}$
(dashed) of accepted cluster and field stars, respectively, as
function of $\epsilon$ for the specific configuration with cluster
centre coordinates $(\ell, b) = (300^\circ\!, 0^\circ\!)$ and distance
$D = 150$ pc. The value $\epsilon_{\rm min} = 0.954$ is indicated by
the dashed vertical line. The two lowest $f_{\rm mg}$-lines correspond
to the two `observations' outside the central band in
Figure~\ref{fig4}.}
\label{fig3}
\end{figure}

%\section 3.2
\subsection{Application to synthetic data}
\label{subsubsec_appl}

We have constructed 20 random realizations of a synthetic cluster
superimposed on a field star distribution for each of the 24 specific
configurations (\S \ref{subsubsec_constr} and
Table~\ref{tab_synth_data}). For each of these 480 synthetic data
sets, the cluster convergent point and members were determined using
the refurbished convergent point method (\S \ref{subsec_mod}). After
extensive experiments, we choose to fix $t_{\rm min} = 1.7$
(eqs~\ref{eq_def_tpm} and \ref{eq_def_tpmnew}), and to use for
$\sigma_{\rm int}$ (eq.~\ref{eq_def_sigma}) the value which was
actually used in the construction of the synthetic data
(2.0~km~s$^{-1}$). The latter choice is motivated by the fact that the
results of the membership selection turned out to be rather
insensitive to the precise value of $\sigma_{\rm int}$ (cf.\ \S
\ref{subsec_appl_open}). Our choice for $t_{\rm min}$ is motivated by
the median distance of the associations discussed by de Zeeuw et al.\
(1999) of $\sim$300~pc. This value combined with $\sigma_{\rm int} =
2.0$~km~s$^{-1}$ gives $\sigma_{\rm int}^\star =
1.4$~mas~yr$^{-1}$. For $\sigma_\mu = 1$~mas~yr$^{-1}$,
eqs~(\ref{eq_def_tpm}) and (\ref{eq_def_tpmnew}) with $t_{\rm min} =
1.7$ imply that all stars with $\mu \leq 3$~mas~yr$^{-1}$ are
rejected, which is identical to rejecting all stars for which the
proper motion is less than the familiar 3$\sigma$. The remaining free
parameter $\epsilon_{\rm min}$ (step~[viii] in \S \ref{subsec_clas})
represents the stop criterion in the selection process. We determine
its optimum value based on Monte Carlo simulations.

%\section 3.2.1
\subsubsection{Effect of $\ell$ and $D$}
\label{subsubsubsec_field_eff1}

\begin{figure}
\centerline{\psfig{file=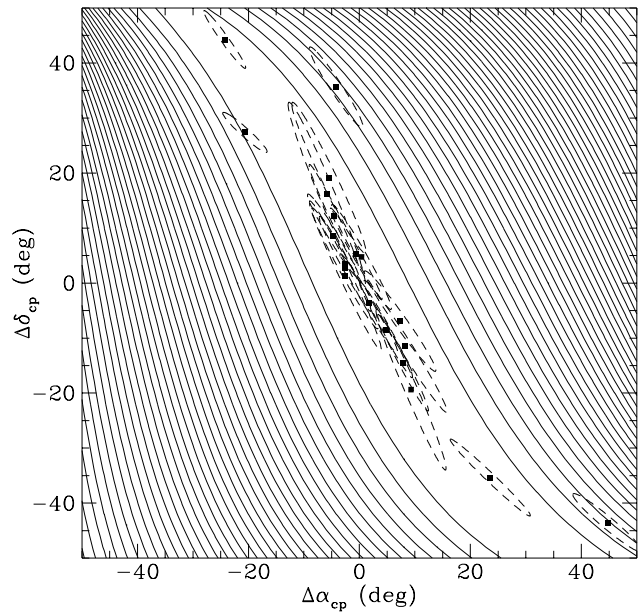,width=8.35cm,silent=}}
\caption{Filled squares indicate the differences between the positions
of the real and determined convergent points for the 20 Monte Carlo
realizations with $(\ell, b) = (300^\circ\!, 0^\circ\!)$ and $D = 150$
pc. The dashed ellipses denote the 3$\sigma$ uncertainty regions of
the determined convergent points (\S \ref{subsubsec_mod3}). The
contours indicate constant values of $X^2$, linearly spaced by $\Delta
X^2 = 2000.00$ with a minimum value $X^2 = 2.19$ at
$(\Delta\alpha_{\rm cp}, \Delta\delta_{\rm cp}) = (0^\circ\!,
0^\circ\!)$ or $(\alpha_{\rm cp}, \delta_{\rm cp}) = (73\fdg897,
-1\fdg747)$, for the `perfect cluster' with Solar motion and Galactic
rotation but without streaming motion. The two `observations' outside
the central band correspond to the two lowest $f_{\rm mg}$-lines in
Figure~\ref{fig3}.}
\label{fig4}
\end{figure}

The convergent point method kinematically detects clusters using
proper motions only. A detection requires the group to stand out from
the field star distribution sufficiently clearly. Field stars are
background and foreground stars as well as non-cluster members at
approximately the cluster distance.

The distinction between background and foreground stars and cluster
members is simple for nearby clusters. This is illustrated in
Figure~\ref{fig1}, which shows the relation between Galactic longitude
$\ell$ (at $b = 0^\circ\!$) and proper motion $\mu$ due to the
distance-independent effect of Galactic rotation (Feast \& Whitelock
1997) and the distance-dependent reflex of the Solar motion (Dehnen \&
Binney 1998) for an object at $D = 150, 300, 450$, and 600~pc. The
figure shows that (1) for given $\ell$ the measured proper motion of a
star is an indicator of its distance, and (2) this indicator is more
sensitive at smaller distances, especially near $\ell = 150^\circ\!$
and $300^\circ\!$.

The cluster streaming motion enables to distinguish between field
stars at the cluster distance and genuine cluster members. In the
simulations, this streaming motion (item 1 in \S
\ref{subsubsubsec_mg_kinematics}) has a maximum value of $10~{\rm
km}~{\rm s}^{-1}$. In the isotropic case, the expectation value for
the tangential streaming velocity equals $15 {\hat \pi} / 8 =
5.89~{\rm km}~{\rm s}^{-1}$ (${\hat \pi} = 3.14\ldots$), which
translates to an amplitude in proper motion of $\mu = 5.89~\pi / A =
1.24~\pi~{\rm mas}~{\rm yr}^{-1}$; this gives $\mu \sim 8.3~{\rm
mas}~{\rm yr}^{-1}$ for $D = 150$~pc, but only $\mu \sim 2.1~{\rm
mas}~{\rm yr}^{-1}$ for $D = 600$~pc. Besides the magnitude of the
streaming motion, its orientation is also important. A `mis-aligned'
streaming motion could cause the proper motions of the group members
to vanish, or could cause them to mix indistinguishably with the
Galactic disk distribution, reducing the astrometric cluster detection
probability.

%\section 3.2.2
\subsubsection{Effect of $\epsilon$}
\label{subsubsubsec_field_eff2}

All results of the membership analysis and convergent point
determination for the 24 specific configurations can be understood
completely in terms of the arguments presented above. As an example,
Figure~\ref{fig2} shows the results for the 20 synthetic data sets
with the cluster centred at $(\ell, b) = (300^\circ\!, 0^\circ\!)$ for
$D = 150, 300, 450$, and 600~pc. The figure shows the selected number
of field stars (20 dashed lines) and cluster stars (20 solid lines) as
function of $\log\epsilon$, the stop parameter. As $\epsilon$
increases, the cluster detection significance level increases
(eq.~\ref{eq_def_eps}), which means that more stars have been rejected
in order to reach the lower $X^2$.  Ideally, this rejection concerns
field stars only. This would lead to solid lines which are constant at
the number of cluster members down to $\epsilon = 1$, and to
decreasing dashed lines, vanishing at $\epsilon = 1$.

\begin{table*}
\caption{The number of selected Hyades members by P98 (Perryman et
al.\ 1998), B99 (this study; $\sigma_{\rm int} = 2.0~{\rm km}~{\rm
s}^{-1}$), B99 {\it and$\,$} P98 (B99 I P98; I for intersection), P98
and {\it not$\,$} B99 (P98 M B99; M for minus), and B99 and {\it
not$\,$} P98 (B99 M P98), for four different regions around the Hyades
cluster centre ($0 \leq r$ [All], $r \leq 10~{\rm pc}$ [`core' +
`corona'], $10 < r \leq 20~{\rm pc}$ [`halo'], and $r > 20~{\rm pc}$
[`moving group population']). The comparison between B99 and P98 is
made for three different sets of P98 members: `3$\sigma$' members are
all stars for which the space velocity and the Hyades centre-of-mass
motion agree to within the 99.73~per cent confidence region; for
`2$\sigma$' members these quantities agree to within 95.4~per cent,
while for `1$\sigma$' members they agree to within 68.3~per cent.}
\label{tab_hyades}
\begin{tabular}{lrrrrrrrrrrrr}
\hline
                              &
\multicolumn{4}{c}{\underline{\null\hskip1.7truecm`3$\sigma$' P98\null\hskip1.7truecm}} &
\multicolumn{4}{c}{\underline{\null\hskip1.7truecm`2$\sigma$' P98\null\hskip1.7truecm}} &
\multicolumn{4}{c}{\underline{\null\hskip1.7truecm`1$\sigma$' P98\null\hskip1.7truecm}}\\
                              &
All   & $r \leq 10$ & $10 < r \leq$ & $r > 20$ &
All   & $r \leq 10$ & $10 < r \leq$ & $r > 20$ &
All   & $r \leq 10$ & $10 < r \leq$ & $r > 20$ \\
\medskip
                              &
      & pc          & $\leq 20$~pc  & pc       &
      & pc          & $\leq 20$~pc  & pc       &
      & pc          & $\leq 20$~pc  & pc       \\
P98      &  218& 134& 45&  39&  190& 131&  38&  21&  162& 121&  29&  12\\
B99      &  290& 138& 49& 103&  290& 138&  49& 103&  290& 138&  49& 103\\
B99 I P98&  203& 133& 42&  28&  186& 131&  37&  18&  161& 121&  29&  11\\
P98 M B99&  15 &   1&  3&  11&    4&   0&   1&   3&    1&   0&   0&   1\\
B99 M P98&  87 &   5&  7&  75&  104&   7&  12&  85&  129&  17&  20&  92\\
\hline\\ 
\end{tabular} 
\end{table*}

Figure~\ref{fig2} shows that one generally starts with fewer cluster
members than $N_{\rm mg}$ because some of them are rejected due to
their insignificant proper motions (eqs~\ref{eq_def_tpm} and
\ref{eq_def_tpmnew}). Furthermore, the number of selected field stars
does not vanish at $\epsilon = 1$. This indicates that the membership
lists are contaminated by field stars (`interlopers'). These stars
have proper motions which are consistent with the projected space
motion of the cluster. Elimination of interlopers requires radial
velocities, photometric data, and/or distance information. The number
of selected cluster stars is not constant with $\epsilon$ but
decreases due to the erroneous rejection of members, caused by, e.g.,
velocity dispersion, observational errors, and/or an unfavorable
streaming velocity. Especially for large cluster distances, many
members have sometimes been rejected before an acceptable solution is
reached: in these cases, the combined reflex of the streaming motion,
Galactic rotation, and Solar motion is insufficient to kinematically
detect the cluster using proper motions only.

Figure~\ref{fig3} shows, for $(\ell, b) = (300^\circ\!, 0^\circ\!)$
and $D = 150$~pc, the fractions $f_{\rm mg}$ and $f_{\rm field}$
of selected cluster and field stars, respectively, as function of
$\epsilon$. The convergent point method confirms $\sim$80~per cent of
the cluster members, but with a contamination of $\sim$20~per cent of
the total number of field stars in the field of view. Simulations show
that these percentages are practically independent of the total number
of cluster or field stars in the simulation.

The optimum value for the stop parameter $\epsilon$, $\epsilon_{\rm
min}$, is the value which returns the largest number of cluster
members with the smallest contamination by field stars. It follows
from Figures~\ref{fig2} and \ref{fig3} that $\epsilon_{\rm min}$
should be chosen close to 1, independent of cluster distance. This
conclusion is supported by results of the other synthetic data
sets. We have chosen $\epsilon_{\rm min} = 0.954$ ($\log\epsilon_{\rm
min} = -0.020$).

%\section 3.2.3
\subsubsection{Convergent point}
\label{subsubsubsec_field_eff3}

Figure~\ref{fig4} compares the positions of the real convergent point
of the cluster and the one determined by the convergent point method
for the 20 Monte Carlo realizations with $(\ell, b) = (300^\circ\!,
0^\circ\!)$ and $D = 150$~pc; we define $\Delta\alpha_{\rm cp} \equiv
\alpha_{\rm cp, obs} - \alpha_{\rm cp}$ and $\Delta\delta_{\rm cp}
\equiv \delta_{\rm cp, obs} - \delta_{\rm cp}$. The `observations'
form an elongated structure, extending over several tens of degrees,
in the $\Delta\alpha_{\rm cp}$--$\Delta\delta_{\rm cp}$-plane.

This effect is explained by the contours in Figure~\ref{fig4}, which
represent constant values of $X^2$ (eq.~\ref{eq_def_chi}).  The
contours correspond to the `perfect cluster' at $(\ell, b) =
(300^\circ\!, 0^\circ\!)$ and $D = 150$~pc: 200 stars with proper
motions due to the reflex of Solar motion and Galactic rotation only.
Each of these proper motion vectors defines a great circle. Because
the proper motions for all stars are nearly parallel, the
corresponding great circles are nearly parallel. As a result, the
proper motions define quite accurately a great circle on which the
convergent point must be located; however, the precise location of the
convergent point on this great circle is rather uncertain. The
contours in Figure~\ref{fig4} show exactly this: the convergent point
has a large freedom, in terms of $X^2$, to move along a great circle
whereas excursions in the direction perpendicular to this great circle
are not allowed (e.g., Bertiau 1958; Perryman et al.\ 1998).

%\section 4
\section{Application}
\label{sec_appl}

%\section 4.1
\subsection{Hyades open cluster}
\label{subsec_appl_open}

Perryman et al.\ (1998; P98) combined Hipparcos proper motions and
parallaxes with ground-based radial velocities in order to establish
membership for the Hyades open cluster. Their sample contains all
stars in the Hipparcos Catalogue in the field $2^{\rm h}\,15^{\rm m} <
\alpha < 6^{\rm h}\,5^{\rm m}$ and $-2^\circ\! < \delta < 35^\circ\!$
with $\pi \geq 10~{\rm mas}$. P98 define as members all stars for
which the space velocity coincides with the Hyades centre-of-mass
motion to within the 99.73~per cent confidence region
(`3$\sigma$'). For stars lacking radial velocity information, the
decision is based on tangential velocity only. P98 start with 142
secure pre-Hipparcos members in the central region of the
Hyades. Convergence of their procedure was achieved after two
iterations, resulting in 218 members, 179 of which were previously
known (all with known radial velocities), and 39 of which are new (18
of these have a known radial velocity). The P98 member selection is
generous; only few genuine Hyades members have probably not been
selected, whereas some field stars are likely to be present in the
member list.

\begin{figure*}
\centerline{\psfig{file=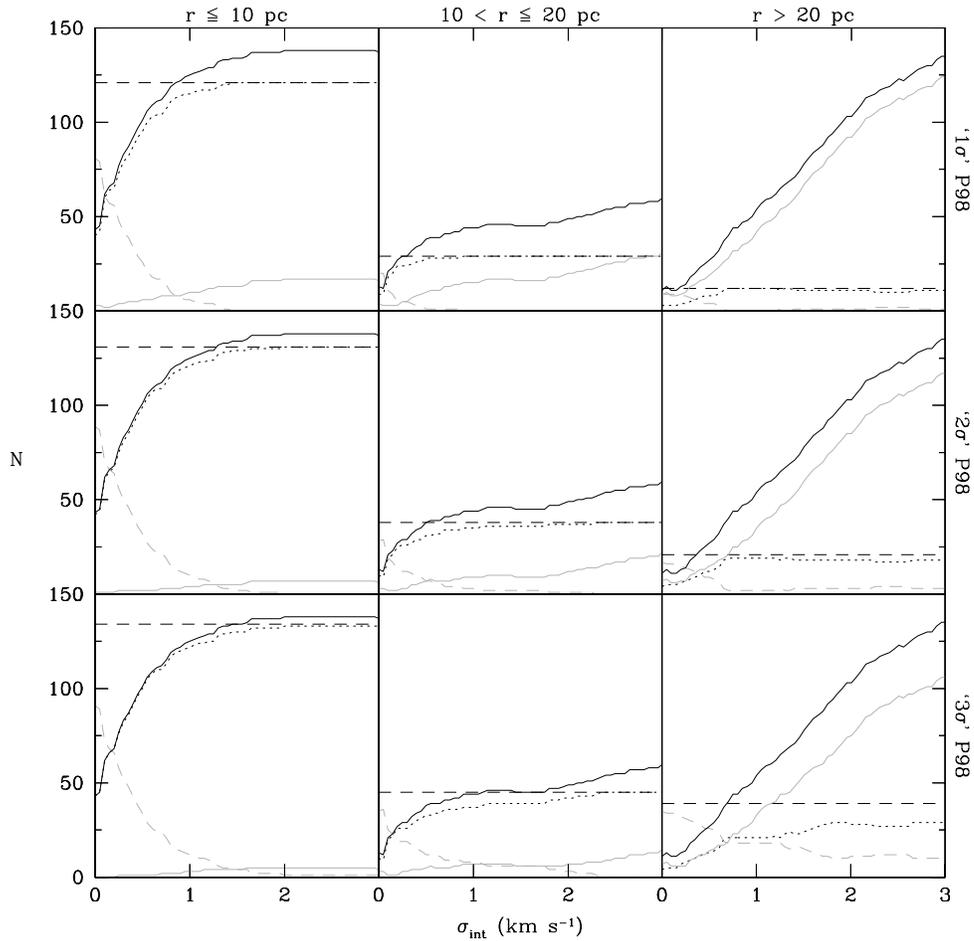,width=13.0truecm,silent=}}
\caption{The number of Hyades members found by B99 (this study; solid
black), by P98 (dashed black), the intersection B99 I P98 (dotted
black), the difference P98 M B99 (dashed grey), and the difference B99
M P98 (solid grey), as function of the (one-dimensional) velocity
dispersion $\sigma_{\rm int}$ used in the selection procedure. From
the top row down: `1$\sigma$' P98 results (confidence limit $p_{\rm
conf} = 0.683$), `2$\sigma$' P98 results ($p_{\rm conf} = 0.954$), and
`3$\sigma$' P98 results ($p_{\rm conf} = 0.9973$). From the left
column right: stars with $r \leq 10$~pc (`core' + `corona'), stars
with $10 < r \leq 20$~pc (`halo'), and stars with $r > 20$~pc (`moving
group population').}
\label{fig5}
\end{figure*}

\begin{figure*}
\centerline{\psfig{file=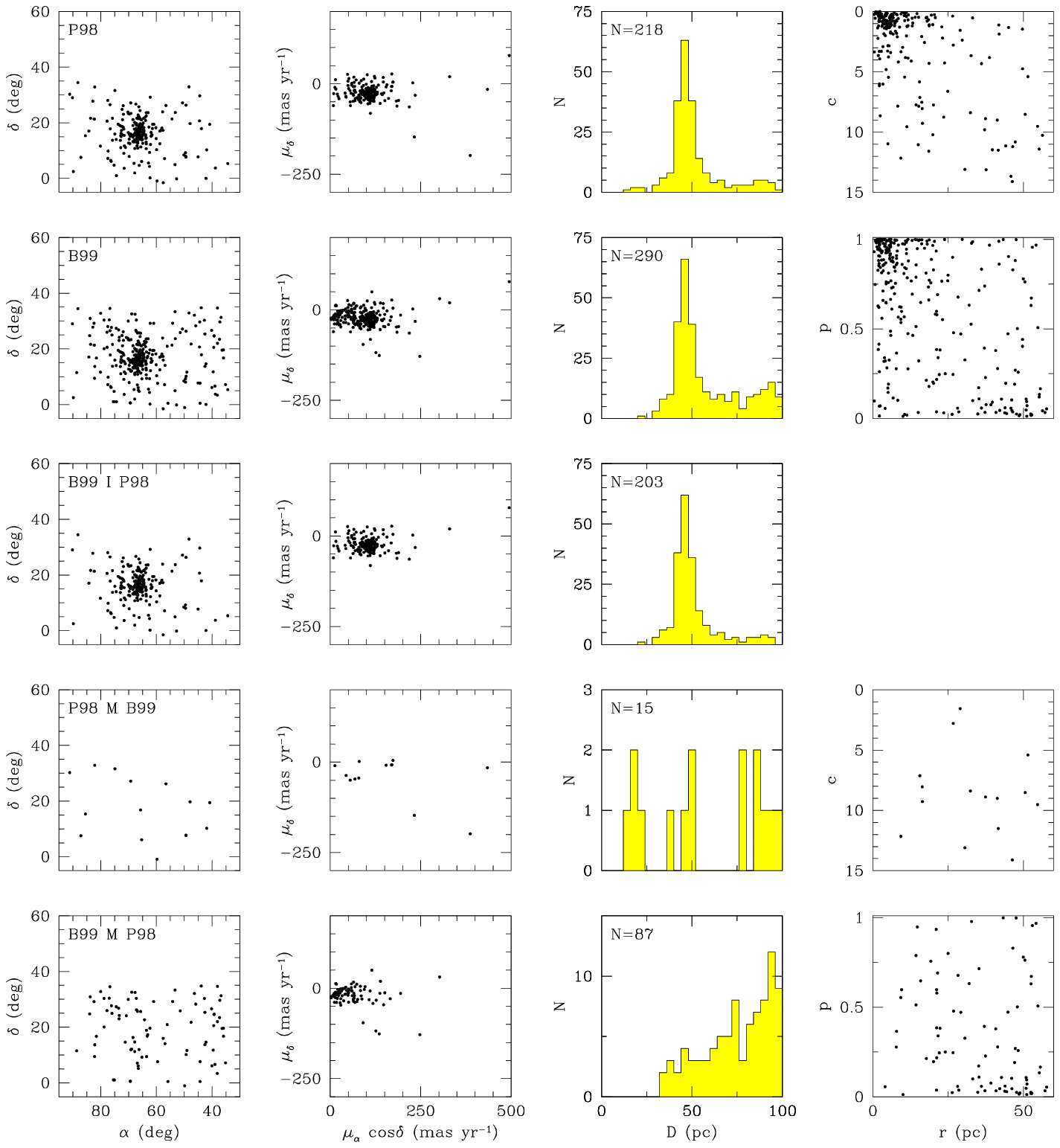,width=16.5truecm,silent=}}
\caption{Results of the membership selection of the Hyades by P98
(`3$\sigma$' members; top row), B99 (this study; $\sigma_{\rm int} =
2.0~{\rm km}~{\rm s}^{-1}$; second row), B99 {\it and$\,$} P98 (B99 I
P98; third row), P98 and {\it not$\,$} B99 (P98 M B99; fourth row),
and B99 and {\it not$\,$} P98 (B99 M P98; fifth row). From the left
column right: (1) positions on the sky; (2) vector point diagram
$(\mu_\alpha \cos\delta, \mu_\delta)$; (3) distance histogram ($D
\equiv 1000 / \pi$; $D$ in pc, $\pi$ in mas) with number of stars
indicated; and (4) membership statistic or probability as function of
three-dimensional distance $r$ (pc) from the Hyades cluster centre:
$p$ ($0 \leq p \leq 1$; eq.~\ref{eq_def_pnew}) versus $r$ for B99 and
B99 M P98; the dimensionless quantity $c$, defined by $c \equiv
\bmath{z}^{\rm T} {\bf \Sigma}^{-1} \bmath{z}$, where T denotes taking
the transpose, $\bmath{z}$ is the difference vector between the
observed space velocity of the star and the Hyades centre-of-mass
motion, and ${\bf \Sigma}$ is the sum of the two corresponding
covariance matrices, for P98 and P98 M B99. In case $\bmath{z}$ is
$\nu$-dimensional, $c$ is distributed as $\chi^2$ with $\nu$ degrees
of freedom: for $\nu = 2$ (no radial velocity information), $c =
11.83$ corresponds to a confidence limit $p_{\rm conf} = 0.9973$
(`3$\sigma$'), and for $\nu = 3$, $c=14.16$ relates to $p_{\rm conf} =
0.9973$ (`3$\sigma$').}
\label{fig6}
\end{figure*}

Figure~8(a) in P98 shows the positional distribution of the 218
members. P98 consider four distinct components ($r$ is defined as the
three-dimensional distance to the cluster centre): (1) a spherical
`core' with a core radius $r_{\rm c} \sim 2.7~{\rm pc}$ and half-mass
radius $r_{\rm h} \sim 5.7~{\rm pc}$; (2) a `corona' extending out to
the tidal radius $r_{\rm t} \sim 10~{\rm pc}$; (3) a `halo' consisting
of stars with $r_{\rm t} \la r \la 2 r_{\rm t}$ which are still
dynamically bound to the cluster; and (4) a `moving group population'
of stars with $r \ga 2 r_{\rm t}$ which have similar kinematics as the
central parts. The halo and the moving group population are explained
by evaporation and diffusion of core and corona members due to the
effects of the Galactic tidal field and stellar encounters (\S
\ref{subsec_appl_super}). The core and corona contain 134 stars ($r
\leq 10~{\rm pc}$), the halo contains 45 stars ($10 < r \leq 20~{\rm
pc}$), and the moving group population contains 39 stars ($r > 20~{\rm
pc}$). Kinematic assignment of membership beyond $\sim$$2 r_{\rm t}$
is difficult, and thus less reliable.

Figure~8(b) in P98 shows the velocity distribution of the 218
members. The expected one-dimensional velocity dispersion is
0.2--$0.4~{\rm km}~{\rm s}^{-1}$, too low to be resolved: the observed
velocities are fully consistent with parallel space motions of the
members, with a (one-dimensional) velocity dispersion of
$\sim$$0.3~{\rm km}~{\rm s}^{-1}$ due to a combination of internal
motions, undetected binaries, and measurement errors.

We apply the convergent point method to the Hyades with $D = 45~{\rm
pc}$, $t_{\rm min} = 1.7$, and $\epsilon_{\rm min} = 0.954$.
Figure~\ref{fig5} shows the results as function of the
(one-dimensional) velocity dispersion $\sigma_{\rm int}$ used in the
selection. It follows from Figure~\ref{fig5} that: (1) in order to
reproduce the P98 results, a non-zero value for $\sigma_{\rm int}$ is
required; for $\sigma_{\rm int} \ga 1.5~{\rm km}~{\rm s}^{-1}$, the
numbers of selected stars are comparable to the numbers found by P98;
(2) despite the fact that P98 have made use of additional parallax and
radial velocity data, the convergent point method is able to retrieve
nearly all P98 members in the inner 20~pc of the cluster centre,
without introducing many spurious members at the same time. Most of
the stars that are not picked up by the convergent point method have a
low P98 membership probability (cf.\ Figure~\ref{fig6}).  The
convergent point method picks up a number of field stars in the outer
regions of the cluster, $r \ga 2 r_{\rm t} \sim 20~{\rm pc}$ (cf.\
Figure~\ref{fig6}). Parallaxes and/or radial velocities are required
in order to reject these stars as candidate Hyades members.

Table~\ref{tab_hyades} gives the comparison of this study (B99) and
P98 results for $\sigma_{\rm int} = 2.0~{\rm km}~{\rm
s}^{-1}$. Figure~\ref{fig6} shows the results of the membership
selection and the comparison with the `3$\sigma$' P98 results for
$\sigma_{\rm int} = 2.0~{\rm km}~{\rm s}^{-1}$. The convergent point
method correctly identifies 203 of the 218 P98 Hyades members. Only 15
P98 Hyades members are not picked up by the convergent point method: 4
of the 15 lie between 9.30 and 16.47~pc from the cluster centre, but
have a low membership probability. The same is true for the remaining
11 stars, at $r \geq 26.70~{\rm pc}$. Furthermore, only 4 of them have
a known radial velocity. Eighty-seven new candidate Hyades members are
identified by the convergent point method. Five of them have $r \leq
10~{\rm pc}$, but rather low membership probabilities ($p \leq
0.60$). Seven stars have $10 < r \leq 20~{\rm pc}$. The remaining 75
candidates have $r > 20~{\rm pc}$, combined with low membership
probabilities as well as predominantly small proper motions and
parallaxes ($D \ga 50~{\rm pc}$), and are distributed uniformly over
the sky. This indicates that these stars are most likely field stars,
which can be rejected as candidate members using parallax and/or
radial velocity information.

We find $X^2 = 245.574$ ($0.853$ per degree of freedom), and
$(\alpha_{\rm cp}, \delta_{\rm cp}) = (95\fdg54 \pm 0\fdg43, 7\fdg28
\pm 0\fdg19)$ with a correlation coefficient $\rho = -0.79$ on the
measurement errors (\S \ref{subsubsec_mod3}). Projecting the Hyades
centre-of-mass motion (table~3 in P98) gives corresponding convergent
points of $(\alpha_{\rm cp}, \delta_{\rm cp}) = (97\fdg91, 6\fdg66)$
for the 134 stars with $r \leq 10~{\rm pc}$, and $(97\fdg96, 6\fdg61)$
for the 179 stars\footnote{
We neglect the small difference in cluster centre coordinates between
the 134 stars within 10~pc and the 180 stars within 20~pc of the
cluster centre (table~3 in P98), and correspondingly only have 179
stars within 20~pc of the cluster centre.}
with $r \leq 20~{\rm pc}$. When we apply the convergent point method
to the 218 P98 members as starting set, we end up with 213 members
having $X^2 = 151.712$ ($0.702$ per degree of freedom), $(\alpha_{\rm
cp}, \delta_{\rm cp}) = (97\fdg81 \pm 0\fdg52, 6\fdg74 \pm 0\fdg21)$,
and $\rho = -0.84$. The 5 rejected stars have $r > 30~{\rm pc}$, and 4
of them have no radial velocity information. The different convergent
points seem to show a large spread. However, it can be explained fully
by the errors: the confidence ellipses are elongated along the great
circle which connects the convergent points and the Hyades cluster
centre, and the spread in the convergent points is exactly along this
great circle. Accordingly, the position of the convergent point is
well-defined in the direction perpendicular to this great circle,
whereas it can easily shift along the great circle (\S
\ref{subsubsubsec_field_eff3}). This is illustrated by the rejection
of the last star in the application of the convergent point method to
the 218 P98 members; this step shifts the convergent point from
$(\alpha_{\rm cp}, \delta_{\rm cp}) = (96\fdg70, 6\fdg91)$ to
$(97\fdg81, 6\fdg74)$.

%\section 4.2
\subsection{Hyades supercluster}
\label{subsec_appl_super}

\begin{figure}
\centerline{\psfig{file=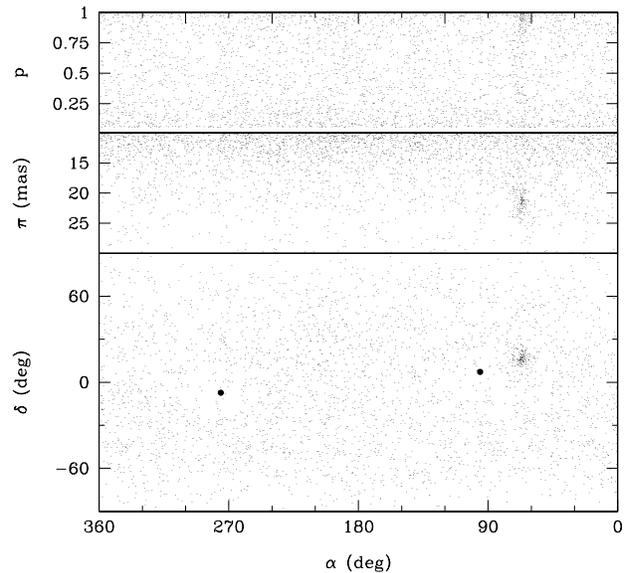,width=8.35truecm,silent=}}
\caption{The 3392 `comoving' Hyades members with $\pi \geq 10$~mas:
sky distribution (bottom), parallax distribution (middle), and
membership probability distribution (top). The two solid dots in the
bottom panel denote the convergent point $(\alpha_{\rm cp},
\delta_{\rm cp}) = (95\fdg54, 7\fdg28)$ and its mirror point
$(275\fdg54, -7\fdg28)$.}
\label{fig7}
\end{figure}

Although it has often been claimed that the Hyades open cluster is
related to a kinematically similar `supercluster' (e.g., Hertzsprung
1909; Eggen 1992; Chereul, Cr\'ez\'e \& Bienaym\'e 1998), these
results are not commonly accepted (e.g, Ratnatunga 1988; Skuljan,
Cottrell \& Hearnshaw 1997). This extended moving group population is
generally explained by the evaporation of the Hyades open cluster
through (1) relaxation by internal stellar encounters, (2) the
Galactic tidal field, and (3) passing interstellar clouds (e.g.,
Terlevich 1987). Although all these effects are indisputably taking
place, it is not to be expected that all escaped members have
velocities consistent with the cluster space motion: stars beyond the
cluster tidal radius are subject to Galactic tidal forces and to the
shearing effect of differential Galactic rotation, and therefore can
have systematically distorted motions. A selection criterion different
from velocity, e.g., angular momentum, might therefore be more
appropriate to identify escaped members (e.g., Innanen, Harris \&
Webbink 1983; Helmi, Zhao \& de Zeeuw 1998; P98). In short,
establishing the significance of the reality and the interpretation of
superclusters and all-sky stellar streams is an important issue
requiring a thorough analysis, beyond the scope of this paper.

Nonetheless, an interesting question is: `is there a significant
over-abundance of low-$t_\perp$ stars for the Hyades convergent point
over the whole sky?' We have applied the convergent point method to
the entire Hipparcos Catalogue, subject to the constraint $\pi \geq
10$~mas, to search for `comoving' Hyades members. Out of 22951
targets, we find 3392 such stars, the overwhelming majority of which
(94~per cent) has $r > 20$~pc. However, their distribution on the sky
is completely random, their parallax distribution peaks sharply
towards $\pi = 10$~mas, and their membership
probabilities are distributed uniformly (Figure~\ref{fig7}).  All
these facts are indicative of a non-physical grouping. Indeed,
searching for all-sky comoving stars at random cluster velocities
(convergent points) yields identical results. An all-sky search using
the ACT/TRC catalogues (\S \ref{sec_disc}) is practically feasible,
but is prone to give even less well-interpretable results because
these catalogues (1) each contain 10 times more stars, (2) have proper
motion accuracies which are 2--3 times worse than those of Hipparcos,
and (3) do not allow the use of parallaxes to make the $\pi \geq
10$~mas cutoff (giving in practice a 50 times larger
sample). Obviously, proper motion data alone is insufficient for
all-sky moving group analyses; an in-depth investigation of multi-band
photometric, spectroscopic, and/or parallax data is required to bring
down the number of interlopers significantly.

%\section 5
\section{Discussion}
\label{sec_disc}

The Hipparcos Catalogue (ESA 1997) has opened a new era of stellar
kinematical research. The high quality of the data, its homogeneity,
and the absence of systematic errors down to the 0.1~mas level, make
the Hipparcos proper motions ideally suited for a reassessment of the
existence and membership of moving groups in the Solar
neighbourhood. As Hipparcos parallaxes do not yield useful constraints
on individual tangential space velocities for stars at distances $D
\ga 350$--500~pc, a common motion of distant stars is best detected by
converging proper motions. The nature of the Hipparcos data,
specifically the full covariance matrices provided in the Catalogue
and its high quality which sometimes allows to marginally resolve
internal motions, combined with increased computer power over the last
decades, have necessitated a refurbishment of the convergent point
method that was originally developed by Jones in 1971. This method is
a maximum likelihood procedure which detects moving groups and selects
the corresponding members from a given set of stars with positions and
proper motions. The method selects all stars which have proper motions
that are consistent with the maximum likelihood convergent point, and
assigns individual membership probabilities. The method is known to
work well for nearby groups, such as the Hyades, which span a large
enough area on the sky so as to allow the proper motions to show their
converging pattern (see Cudworth 1998 for a review). The method is
biased towards distant stars, as these have generally small proper
motions. Therefore, such stars are more likely not to be rejected as
members.

Extensive Monte Carlo simulations have been used to fix free
parameters. These simulations show that typically $\sim$80~per cent of
all cluster members can be retrieved, combined with a contamination of
field stars which amounts to $\sim$20~per cent of the total number of
(field) stars in the sample (Figure~\ref{fig3}). We presented an
application of our method to the Hyades with Hipparcos positions and
proper motions. The results for stars near the centre of the cluster
agree with those of Perryman et al.\ (1998), who used parallaxes and
radial velocities as well to determine membership. An a posteriori
inspection of the Hipparcos parallaxes of stars that are selected by
the convergent point method but not by Perryman et al.\ confirms that
nearly all of them lie at large distances from the Hyades cluster
centre (Figure~\ref{fig6}). The majority of these stars have low to
medium membership probabilities, consistent with them being field
stars.

The abovementioned bias can also be removed by combining the
convergent point method with other selection procedures, such as
Hoogerwerf \& Aguilar's (1999) `Spaghetti method' which also uses
parallax information. De Zeeuw et al.\ (1999) have done so, and
successfully applied the combined procedure to nearby OB associations
with Hipparcos data. As expected, the combination of the two methods
efficiently eliminates the biases in the individual methods (figure~4
in de Zeeuw et al.). De Zeeuw et al.\ used a two-step strategy to
determine membership, thereby increasing the sensitivity and power of
the convergent point method. In the first step, they identified the
association using the early-type stars in the Catalogue, with the
projected space motion, i.e., convergent point, as result. As the
number of field stars among the early-type stars is generally several
factors smaller than the numbers assumed in our Monte Carlo
simulations (\S\S
\ref{subsubsubsec_field_eff1}--\ref{subsubsubsec_field_eff3};
Table~\ref{tab_synth_data}), the convergent point method was able to
detect associations out to distances of $\sim$650~pc. In the second
step, additional association members of later spectral type were
selected from the remaining stars in the Catalogue by assuming the
convergent point of the late-type members is equal to that of the
early-type members.

Several parametric and non-parametric methods for finding moving
groups have recently been developed (Chen et al.\ 1997; Figueras et
al.\ 1997; Chereul et al.\ 1998; Hoogerwerf \& Aguilar 1999; see also
Dehnen 1998). All of them use Hipparcos positions, proper motions, and
parallaxes; some of them also require ground-based radial velocities.
The latter methods are currently of limited use, as a uniform set of
homogeneous radial velocities for the majority of stars in the
Hipparcos Catalogue is not available (cf.\ Udry et al.\ 1997); the
lack of reliable radial velocities for early-type stars is especially
severe, so that these methods cannot be used for detection and
membership selection of OB associations.

The positions for $\sim$$10^6$ stars with $V \la 11$~mag in the
Astrographic Catalogue have recently been combined with those in the
Tycho Catalogue. The resulting ACT and TRC catalogues list proper
motions on the Hipparcos reference frame with accuracies
$\sim$3~mas~yr$^{-1}$ (Urban et al.\ 1998; H{\o}g et al.\ 1998).  The
ACT/TRC catalogues are presently not the ideal catalogues to apply the
convergent point method to: in order to bring down the number of
interlopers to an acceptable level, a large amount of additional
spectroscopic and/or multi-band photometric data is indispensable (\S
\ref{subsec_appl_super}). Future astrometric satellite missions such
as GAIA will provide micro-arcsecond astrometry and accurate
multi-epoch multi-band photometry down to $V \sim 20$~mag as well as
radial velocities accurate to $\sim$3~km~s$^{-1}$ down to $V \sim
15$~mag for $\sim$$10^9$ stars (e.g., Perryman, Lindegren \& Turon
1997; Gilmore et al.\ 1998). Such data will allow the detection of
common motion of stars throughout the Galaxy!

%%%%%%%%%%%%%%%
% Acknowledgments
%%%%%%%%%%%%%%%

\section*{Acknowledgments}
It is a pleasure to thank Luis Aguilar, Adriaan Blaauw, Anthony Brown,
Ronnie Hoogerwerf, Michael Perryman, and Tim de Zeeuw for numerous
stimulating discussions and/or a careful reading of the manuscript.
The referee is acknowledged for constructive criticism and helpful
comments.

%%%%%%%%%%%%%%%
% Bibliography
%%%%%%%%%%%%%%%

%%%%%%%%%%%%%%%
% Appendices
%%%%%%%%%%%%%%%

\appendix

%%%%%%%%%%%%%%%
% Appendix A
%%%%%%%%%%%%%%%

%\section A
\section{From $\bmath{(\mu_\alpha \cos\delta, \mu_\delta)}$ to $\bmath{(\mu_\parallel, \mu_\perp)}$}
\label{app_mu_trans}

For a given convergent point $(\alpha_{\rm cp}, \delta_{\rm cp})$, the
transformation of $(\mu_\alpha \cos\delta \equiv \mu_{\alpha^\ast},
\mu_\delta)$ to $(\mu_\parallel, \mu_\perp)$ can be carried out using
the general recipe outlined in ESA (1997, Vol.\ 1, \S 1.5.2). The
transformation uses a 4-dimensional vector $\bmath{a}$:
\begin{equation}
\bmath{a} \equiv
\left(\begin{array}{c}
\alpha^\ast       \\
\delta            \\
\mu_{\alpha^\ast} \\
\mu_\delta
\end{array}\right)
\stackrel{f}{\qquad\longrightarrow\qquad}
\left(\begin{array}{c}
\alpha^\ast   \\
\delta        \\
\mu_\parallel \\
\mu_\perp
\end{array}\right)
\equiv \hat{\bmath{a}},
\end{equation}
where $\alpha^\ast \equiv \alpha \cos\delta$. The covariance matrix of
$\hat{\bmath{a}}$, ${\bf C}_{\hat{\bmath{a}}}$, is given by ${\bf
C}_{\hat{\bmath{a}}} \equiv {\bf J}_{f}~{\bf C}_{\bmath{a}}~{\bf
J}_{f}^{\rm T}$, where T denotes taking the transpose, ${\bf J}_{f}$
denotes the Jacobian matrix of the transformation $f$, and ${\bf
C}_{\bmath{a}}$ is the covariance matrix of $\bmath{a}$. Using
eqs~(\ref{eq_def_perpar}) and (\ref{eq_def_theta}), we find that ${\bf
J}_{f} \equiv \partial(\alpha^\ast, \delta, \mu_\parallel, \mu_\perp)
/ \partial(\alpha^\ast, \delta, \mu_{\alpha^\ast}, \mu_\delta)$ is
given by:
\begin{equation}
\left( \begin{array}{rrrr}
1 & 0 & 0 & 0 \\
0 & 1 & 0 & 0 \\
-\mu_{\perp    } {\rm d}\theta / {\rm d}\alpha^\ast &
-\mu_{\perp    } {\rm d}\theta / {\rm d}\delta      &
 \sin\theta & \cos\theta \\
 \mu_{\parallel} {\rm d}\theta / {\rm d}\alpha^\ast &
 \mu_{\parallel} {\rm d}\theta / {\rm d}\delta      &
-\cos\theta & \sin\theta
\end{array} \right),
\end{equation}
with
\begin{eqnarray}
{{{\rm d}\theta} \over {{\rm d}\alpha^\ast}} & = &
{{\sin^2\theta} \over {\sin^2(\alpha_{\rm cp} - \alpha)}}
  \left(\tan\delta -
        \tan\delta_{\rm cp}\cos[\alpha_{\rm cp} - \alpha]
  \right), \nonumber\\
{{{\rm d}\theta} \over {{\rm d}\delta}} & = &
\alpha\sin\delta \left({{{\rm d}\theta}\over{{\rm d}\alpha^\ast}}\right) +
{{\cos\delta \sin^2\theta} \over {\sin(\alpha_{\rm cp} - \alpha)}} \nonumber\\
& &
  \left(\tan\delta \tan\delta_{\rm cp} +
        \cos[\alpha_{\rm cp}-\alpha]\right).
\end{eqnarray}

%%%%%%%%%%%%%%%
% Appendix B
%%%%%%%%%%%%%%%

%\section B
\section{Membership probability}
\label{app_prob}

The Hipparcos Catalogue provides the five astrometric parameters
$\bmath{a} \equiv (\alpha \cos\delta, \delta, \pi, \mu_\alpha
\cos\delta, \mu_\delta)^{\rm T}$, where T denotes taking the
transpose, together with the 5$\times$5 covariance matrix ${\bf C}$.
Appendix~\ref{app_mu_trans} describes how the proper motion components
$\mu_\alpha \cos\delta$ and $\mu_\delta$ can be transformed into
$\mu_\parallel$ and $\mu_\perp$, with the corresponding 2$\times$2
covariance matrix:
\begin{equation}
{\bf C} = 
\left(\begin{array}{cc}
 \sigma_{\parallel}^2                 & \rho\sigma_{\parallel}\sigma_{\perp} \\
 \rho\sigma_{\parallel}\sigma_{\perp} & \sigma_{\perp}^2
\end{array}\right),
\label{def_eq_C2x2}
\end{equation}
where $\rho$ denotes the correlation coefficient between
$\mu_\parallel$ and $\mu_\perp$. We denote the `measured' proper
motion components $\mu_\parallel$ and $\mu_\perp$ by
$\overline{\mu_\parallel}$ and $\overline{\mu_\perp}$. The confidence
ellipse with confidence limit $p_{\rm conf}$ in the
$\mu_\parallel$--$\mu_\perp$-plane is defined by the locus of points
$(\mu_\parallel, \mu_\perp)$ which solve the equation (Press et
al.\ 1992, p.\ 690--693):
\begin{equation}
\label{eq_conf_ell}
\left( \begin{array}{cc}
   \mu_{\parallel} - \overline{\mu_{\parallel}} &
   \mu_{\perp}     - \overline{\mu_{\perp}}
       \end{array}
\right)
{\bf C}^{-1}
\left( \begin{array}{c}
   \mu_{\parallel} - \overline{\mu_{\parallel}} \\
   \mu_{\perp}     - \overline{\mu_{\perp}}
       \end{array}
\right)
=
\Delta,
\end{equation}
where the parameter $\Delta$ determines the size of the ellipse, and
is related to $p_{\rm conf}$ according to:
\begin{equation}
p_{\rm conf} = {{1}\over{\Gamma\left(\nu / 2 \right)}}
               \int_{0}^{\Delta/2} {\rm d}t\ t^{-1 + \nu/2} \exp{(-t)},
\label{eq_def_p_conf}
\end{equation}
with $\nu$ the number of dimensions. After definition of:
\begin{equation}
\widehat{\mu_{\parallel}}
\equiv
{ {\mu_{\parallel} - \overline{\mu_{\parallel}}}
\over
  {\sigma_{\parallel}}
}
, \qquad \mbox{\rm and} \qquad
\widehat{\mu_{\perp}}
\equiv
{ {\mu_{\perp} - \overline{\mu_{\perp}}}
\over
  {\sigma_{\perp}}
}
,
\end{equation}
eq.~(\ref{eq_conf_ell}) can be written as:
\begin{equation}
\label{eq_conf_ell_2}
       \widehat{\mu_{\parallel}}^2 -
2 \rho \widehat{\mu_{\parallel}} \widehat{\mu_{\perp}} +
       \widehat{\mu_{\perp}}^2
=
\Delta (1 - \rho^2),
\end{equation}
from which follows that:
\begin{equation}
\widehat{\mu_{\perp}}
=
\rho \widehat{\mu_{\parallel}} \pm
\left[
   \left( 1 - \rho^2 \right)
   \left( \Delta - \widehat{\mu_{\parallel}}^2 \right)
\right]^{1/2}.
\end{equation}
A natural membership probability $p$ can be related to the confidence
limit $p_{\rm conf}$ corresponding to the {\it smallest$\,$}
confidence ellipse that is consistent with $\mu_\perp = 0$ in the
point $(\widehat{\mu_\parallel}, \widehat{\mu_\perp}) =
({\widehat{\mu_\parallel^\prime}},
{\widehat{\mu_\perp^\prime}})$. Solving the equations:
\begin{equation}
\left\{ \begin{array}{rcl}
           \widehat{\mu_{\perp^\prime}} & = & - \overline{\mu_{\perp}} /
                                                \sigma_{\perp}, \\
           \left[{\rm d}\widehat{\mu_{\perp}} /
                 {\rm d}\widehat{\mu_{\parallel}}\right]_{
                    \widehat{\mu_{\parallel}} =
                    \widehat{\mu_{\parallel^\prime}}}
              & = & 0,
        \end{array} \right.
\end{equation}
where $\widehat{\mu_\parallel^\prime}$ and
$\widehat{\mu_\perp^\prime}$ are related through
eq.~(\ref{eq_conf_ell_2}), gives the solution
$({\widehat{\mu_\parallel^\prime}}, {\widehat{\mu_\perp^\prime}}) =
(\pm \rho \sqrt{\Delta}, \pm \sqrt{\Delta})$, with the result that:
\begin{equation}
\Delta
=
\left(
   { {\overline{\mu_{\perp}}}
   \over
     {\sigma_{\perp}}
   }
\right)^2 .
\end{equation}
The confidence limit $p_{\rm conf}$ is given by
eq.~(\ref{eq_def_p_conf}) with $\nu = 2$:
$p_{\rm conf} = 1 - \exp{(-\fr{1}{2}\Delta)}$.
Thus, for a given star with measured proper motion components
$\mu_\parallel$ and $\mu_\perp$ and covariance matrix ${\bf C}$ (eq.\
\ref{def_eq_C2x2}) the confidence limit $p_{\rm conf}$ of the {\it
smallest$\,$} confidence ellipse that is consistent with $\mu_\perp =
0$ is given by $p_{\rm conf} = 1 - \exp{(-\fr{1}{2} \left[\mu_\perp
/ \sigma_\perp\right]^2)}$. The membership probability $p$ for this
star is then given by (cf.\ eq.~\ref{eq_def_pconf}):
\begin{equation}
p \equiv 1 - p_{\rm conf} = \exp{(-\fr{1}{2} \left[
         {{\mu_{\perp}} \over {\sigma_{\perp}}} \right]^2)}.
\end{equation}

%%%%%%%%%%%%%%%
% Appendix C
%%%%%%%%%%%%%%%

%\section C
\section{Alternative procedure}
\label{app_alt}

Instead of using $t_\perp$ as membership indicator under the
assumption that its distribution is normal with zero mean and unit
variance, one can also work in $\phi_t$-space, where $\phi_t$ is the
polar angle in the $t$-plane: $t_\parallel \equiv t \sin\phi_t$ and
$t_\perp \equiv t \cos\phi_t$. If $t_\perp$ and $t_\parallel$ are
distributed normally (\S \ref{subsubsec_mod2}) with unit variance and
mean zero and ${\overline t_\parallel}$, respectively, it follows
that:
\begin{eqnarray}
f(\phi_t,t){\rm d}\phi_t{\rm d}t & = & {{{\rm d}\phi_t{\rm d}t}\over{2\pi}}
        t \exp(-\fr{1}{2}[\{t\cos\phi_t\}^2+ \nonumber\\
&&\qquad\qquad\quad        +\{t\sin\phi_t - {\overline t_\parallel}\}^2],
\end{eqnarray}
such that:
\begin{eqnarray}
f(\phi_t){\rm d}\phi_t &\equiv& \int_{0}^{\infty}{\rm d}t f(\phi_t,t)
                                {\rm d}\phi_t =
{{{\rm d}\phi_t}\over{2\sqrt{2\pi}}}
    \exp{\left(-\fr{1}{2}{\overline t_\parallel}^2\right)}\nonumber\\
&& \!\!\!\!\!\!\!\!\!\!\!
    \left(\sqrt{{{2}\over{\pi}}} + a \exp{\left(\fr{1}{2}a^2\right)}
    \left[1 + {\rm erf}\left({{a}\over{\sqrt{2}}}\right)\right]\right),
\end{eqnarray}
where ${\rm erf}(x)$ is the error function (e.g., Abramowitz \& Stegun
1974, eq.~7.1.1) and $a \equiv {\overline t_\parallel}
\sin\phi_t$. For ${\overline t_\parallel} > 0$ $(< 0)$, the function
$f(\phi_t)$ peaks at $\phi_t = \fr{\pi}{2}$ $(\fr{3\pi}{2})$; the
width of the peak increases with decreasing $|{\overline
t_\parallel}|$, i.e., with increasing cluster distance. Thus, an
alternative to minimizing $X^2$ (eq.~\ref{eq_def_chi}) is minimizing:
\begin{equation}
X_{\phi_t}^2 \equiv \sum_{j=1}^{N}\left(\phi_{t,j}-\fr{\pi}{2}\right)^2
\quad{\rm or}\quad
\sum_{j=1}^{N}\left(\phi_{t,j}-\fr{3\pi}{2}\right)^2.
\end{equation}

%%%%%%%%%%%%%%%
% Blackwell reference
%%%%%%%%%%%%%%%

\bsp

\end{document}